\documentclass[preprint,
 amsmath,amssymb,
 aps]{revtex4-1}

\usepackage{graphicx}
\usepackage{dcolumn}
\usepackage{bm}

\begin{document}

\preprint{APS/123-QED}

\title{Expected degree of finite preferential attachment networks}

\author{Michael Small}
\email{michael.small@uwa.edu.au}
\affiliation{School of Mathematics and Statistics\\The University of Western Australia, Crawley, WA, Australia, 6009}

\date{\today}

\begin{abstract}
We provide an analytic expression for the quantity described in the title. Namely, we perform a preferential attachment growth process to generate a scale-free network. At each stage we add a new node with $m$ new links. Let $k$ denote the degree of a node, and $N$ the number of nodes in the network. The degree distribution is assumed to converge to a power-law (for $k\geq m$) of the form $k^{-\gamma}$ and we obtain an exact implicit relationship for $\gamma$, $m$ and $N$. We verify this with  numerical calculations over several orders of magnitude. Although this expression is exact, it provides only an implicit expression for $\gamma(m)$. Nonetheless, we provide a reasonable guess as to the form of this curve and perform curve fitting to estimate the parameters of that curve --- demonstrating excellent agreement between numerical fit, theory, and simulation. \end{abstract}

\pacs{Valid PACS appear here}
\maketitle

\section*{Expected degree}
Preferential attachment \cite{aB99} is the archetypal growth mechanism for scale-free networks. Asymptotically, under certain circumstances, such network produce a degree distribution which converges asymptotically to a power law with exponent $3$. But this is not true in general, and it is not true for arbitrary finite networks generated along the way. In this note we derive straightforward analytic results for the expected exponent $\gamma$ of a scale free network with power law degree distribution $p(k)\propto k^{-\gamma}$. 

We assume that the network is grown with a Barab\'asi-Albert attachment process as described in \cite{aB99}. With each new node we add $m$ links and the growth process is terminated when the network has $N$ nodes. We make the approximation that the degree distribution of this finite networks follows a truncated power-law with some exponent $\gamma$.

Hence, a preferential attachment (PA) network with minimum degree $m$ will add exactly $m$ new links for each new node. The expected degree 
\begin{eqnarray}
\label{spower2}  E(k)&=&2m
\end{eqnarray} 
(since each link has two ends and contributed to the degree of two nodes). Conversely,  the probability that a node has degree $k$ is given by
\begin{eqnarray*}
P(k|\gamma,d) &=&\left\{\begin{array}{cc}
0 & k< m\\
\frac{k^{-\gamma}}{K(\gamma)} & k\geq m
\end{array}
\right.
\end{eqnarray*}
where the normalization factor $K(\gamma)$ is inconvenient. However
\begin{eqnarray*}
\zeta({\gamma} )&=& \left(\sum_{k=1}^{m-1} +\sum_{k=m}^\infty\right)k^{-\gamma} \\
&=&\sum_{k=1}^{m-1}k^{-\gamma}+K(\gamma)
\end{eqnarray*}
and hence it is easily computable.

\begin{figure}
\begin{center}
\begin{tabular}{cc}
\includegraphics[width=0.45\textwidth]{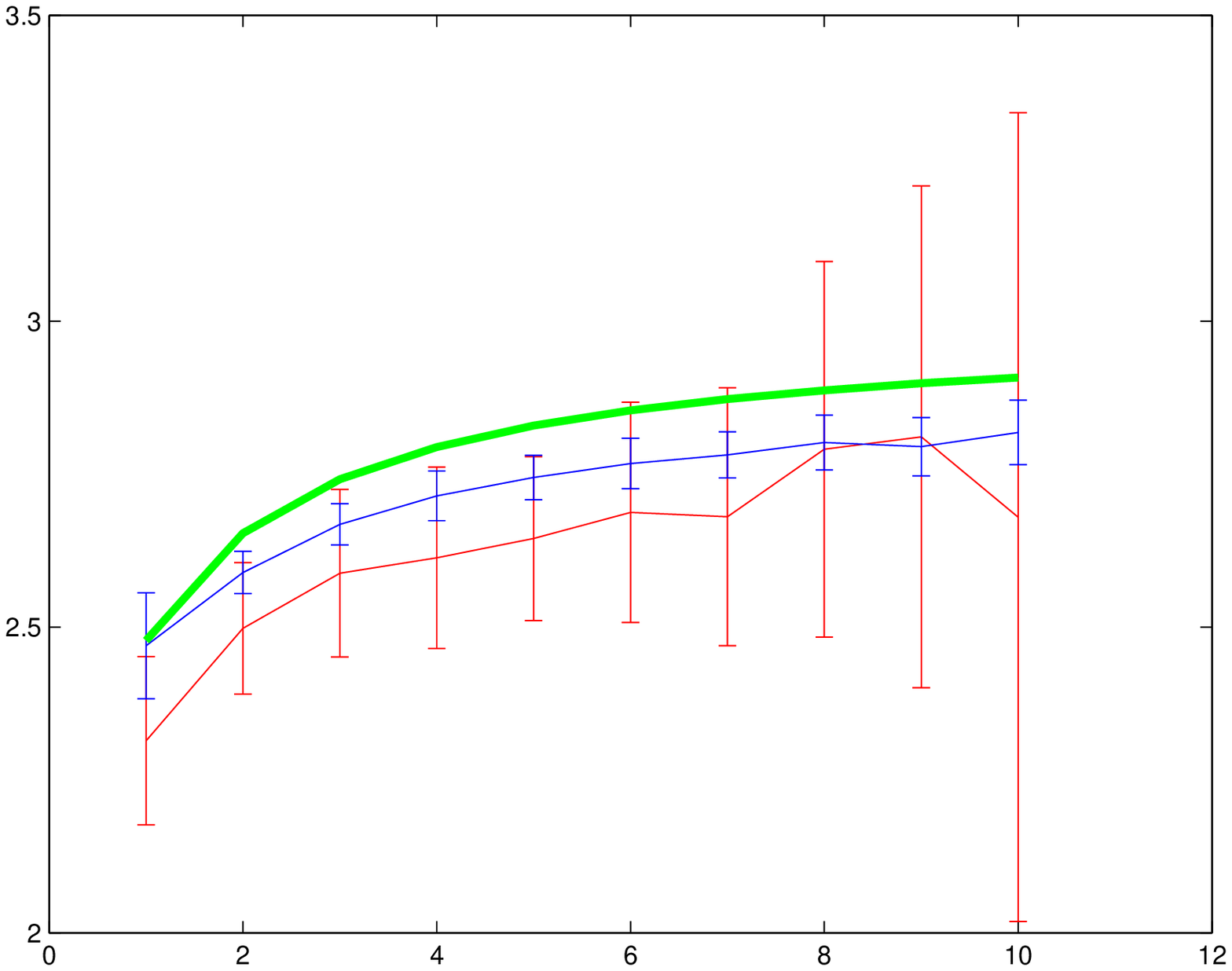} &
\includegraphics[width=0.45\textwidth]{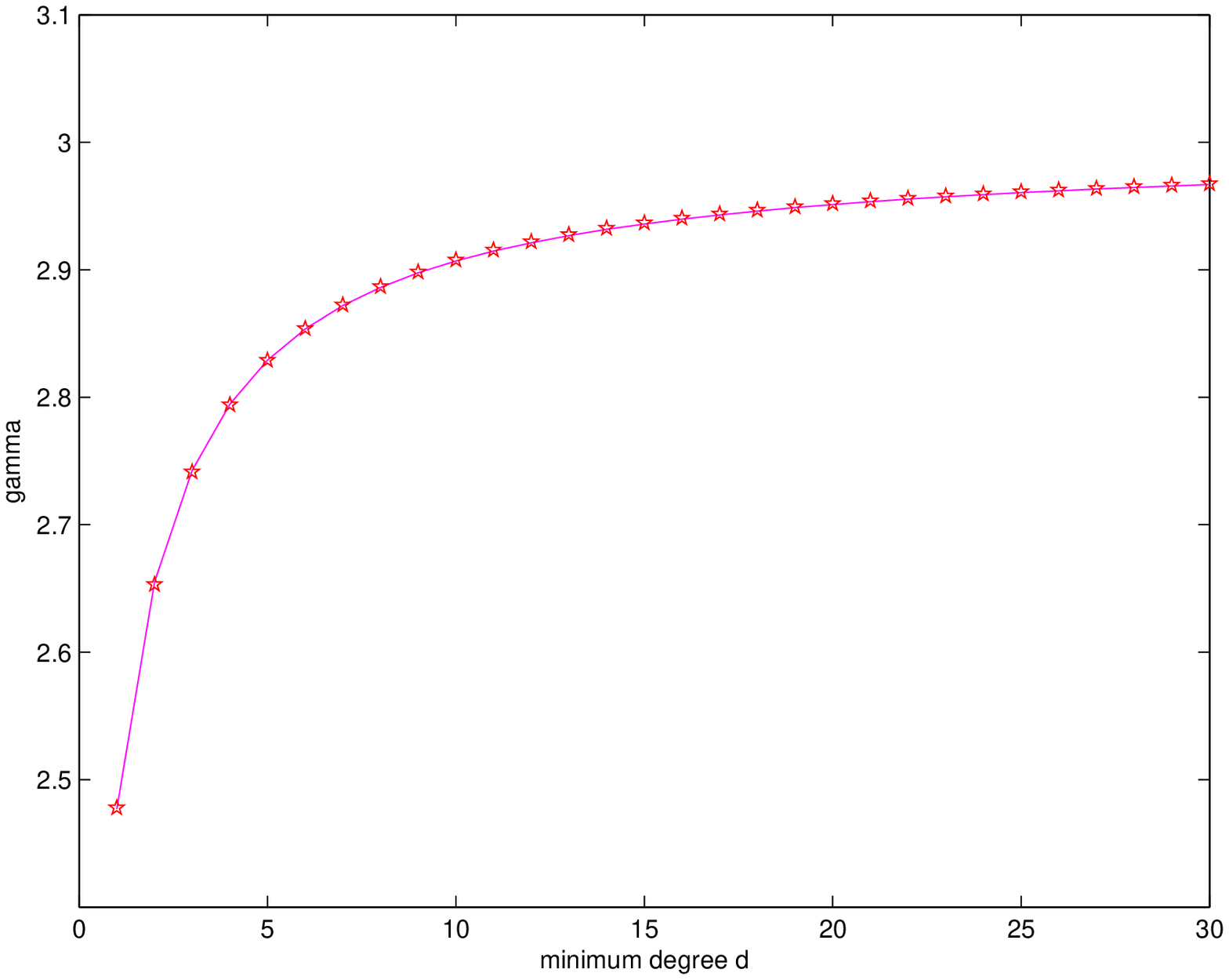}
\end{tabular}
\end{center}
\caption{Left panel: Expected values of $\gamma$ as a function of $m$ (Eqn.  (\ref{egam})) (heavy line) and estimated values of $\gamma$ from $30$ independent realisations of PA networks of size $N$ (mean $\pm$ standard deviation). We take $d\in[1,10]$ and $N=10^3$ (red), $10^4$ (green), $10^5$ (blue). Right panel: $\gamma$ as a function of $m$ computed via the solution of (\ref{mgamm2}) (stars) and estimated from a function fit of the form $\hat\gamma(m)= 3-(m+\alpha)^{-\beta}$. The best fit (obtained from a fit on $m\in[1,10]$) is then extrapolated over the domain. Parameter values are $\alpha= 0.9205$ and $\beta=0.9932$. }
\label{fig_gd}
\end{figure}

The expected degree is
\begin{eqnarray}
\nonumber E(k)  &=&\sum_{k=1}^\infty kP(k|\gamma)\\
\label{egam}
 &=& \frac{\sum_{k=m}^\infty k^{1-\gamma}}{\zeta(\gamma)-\sum_{k=1}^{m-1}k^{-\gamma}}
 \end{eqnarray}
 Equating (\ref{spower2}) and (\ref{egam}), we have that the asymptotic value of $\gamma$ satisfies
 \begin{eqnarray}
 \label{mgamm}
\zeta(\gamma) &=& \sum_{k=1}^{m-1}k^{-\gamma}+\frac{1}{2m}\sum_{k=m}^\infty k^{1-\gamma}.
 \end{eqnarray}
Replacing the RHS of  (\ref{mgamm})  with the corresponding infinite sum and cancelling identical terms we obtain
\begin{eqnarray}
\label{mgamm2}
\sum_{k=m}^\infty (2m-k)k^{-\gamma}&=&0
\end{eqnarray}
 Solving   (\ref{mgamm2})  allows us to determine the expected value of $\gamma $ for the PA algorithm with a particular choice of minimum degree $d$. In particular, for $m=1$ we recover $2\zeta({\gamma})=E(k).$ 
 
 In Fig. \ref{fig_gd} we illustrate the agreement between sample preferential attachment networks of various sizes and the prediction of (\ref{mgamm2}). The curve appears to be asymptotic to $\gamma=3$ and so we fit a function of the form $\hat\gamma(m)= 3-(m+\alpha)^{-\beta}$ to the solution of the series (\ref{mgamm2}). We obtain that
 \[\gamma(m)\approx 3-\frac{1}{(m+0.925)^{0.9932}}.\]
These results are required to explain expected degree distributions observed in a related work \cite{stars}, and in that case also show excellent agreement.

\section*{Acknowledgements}

MS is supported by an Australian Research Council Future Fellowship (FT110100896)

%

\end{document}